\documentclass[aps,prb,twocolumn,superscriptaddress]{revtex4}

\usepackage{graphicx}  
\usepackage{dcolumn}   
\usepackage{bm}        
\usepackage{amssymb}   
\usepackage{amsmath}

\begin{document}


\title{Efficient first-principles calculation of phonon assisted photocurrent in large-scale solar cell devices}


\author{Mattias Palsgaard}
\email[]{mlnp@nanotech.dtu.dk}
\affiliation{Synopsys-QuantumWise A/S, Fruebjergvej 3, Postbox 4, DK-2100 Copenhagen, Denmark}
\affiliation{Department of Micro- and Nanotechnology (DTU Nanotech), Center for Nanostructured Graphene (CNG), Technical University of Denmark, DK-2800 Kgs. Lyngby, Denmark}
\author{Troels Markussen}
\affiliation{Synopsys-QuantumWise A/S, Fruebjergvej 3, Postbox 4, DK-2100 Copenhagen, Denmark}
\author{Tue Gunst}
\affiliation{Department of Micro- and Nanotechnology (DTU Nanotech), Center for Nanostructured Graphene (CNG), Technical University of Denmark, DK-2800 Kgs. Lyngby, Denmark}
\author{Mads Brandbyge}
\affiliation{Department of Micro- and Nanotechnology (DTU Nanotech), Center for Nanostructured Graphene (CNG), Technical University of Denmark, DK-2800 Kgs. Lyngby, Denmark}
\author{Kurt Stokbro}
\affiliation{Synopsys-QuantumWise A/S, Fruebjergvej 3, Postbox 4, DK-2100 Copenhagen, Denmark}

\date{\today}

\begin{abstract}
We present a straightforward and computationally cheap method to obtain the phonon-assisted photocurrent in large-scale devices from first-principles transport calculations. The photocurrent is calculated using nonequilibrium Green's function with light-matter interaction from the first-order Born approximation while electron-phonon coupling (EPC) is included through special thermal displacements (STD).
We apply the method to a silicon solar cell device and demonstrate the impact of including EPC in order to properly describe the current due to the indirect band-to-band transitions.
The first-principles results are successfully compared to experimental measurements of the temperature and light intensity dependence of the open-circuit voltage of a silicon photovoltaic module.
Our calculations illustrate the pivotal role played by EPC in photocurrent modelling to avoid underestimation of the open-circuit voltage, short-circuit current and maximum power.
This work represents a recipe for computational characterization of future photovoltaic devices including the combined effects of light-matter interaction, phonon-assisted tunneling and the device potential at finite bias from the level of first-principles simulations.
\end{abstract}
\maketitle

\section{Introduction}

Photovoltaics (PV) represents a promising technology as a replacement for burning fossil fuels.
In the last couple of decades many promising thin film absorber materials have been discovered, all of them with unique strengths and weaknesses.
CdTe and CIGS (CuInGaSe$_2$) can produce high efficiencies, but include rare and toxic elements, CZTS includes only nontoxic earth abundant elements, but suffers from low efficiency and open circuit voltage ($V_{oc}$)\cite{Fraunhofer2017,Fthenakis2004,Woodhouse2013,Polman2016}.
Clearly there is still room for discovery of new materials to improve on the cost/efficiency relationship.
The field of computational material science has seen massive progression and as a result the difference between system size and complexity attainable in simulations and experiments is becoming smaller every day.
Recently a review was published on the design of new materials using first-principles calculations\cite{Butler2016}. Here it is stressed how the abundance of candidate materials together with the lack of efficient devices highlight the need for efficient predictive device calculations.
Continuum models are used extensively in the field of PV to extract benchmark parameters from measurements on devices and to predict the performance of new device geometries\cite{Burgelman2000}. It is difficult to include important effects such as confinement of electrons and phonons, surface- and strain in the continuum models. These effects can, however, be captured using atomistic models based on density functional theory (DFT). DFT combined with the non equilibrium Green's function (NEGF) formalism was f.ex. previously used to improve a continuum model study of transport through the interface between CZTS and the buffer material CdS important for the CZTS solar cell efficiency\cite{Crovetto2017,Palsgaard2016}. 
In spite of the influx of new thin film based PV cells, silicon remains the market leader and about $90\%$ of PV cells are still based on silicon where large modules with high efficiency and stability can be produced\cite{Fraunhofer2017}. \\
Silicon has an indirect bandgap and as such absorption of a photon around the bandgap energy must be accompanied by the absorption/emission of a phonon to conserve momentum. A number of recent studies also show that EPC plays a key role in the outstanding performance of perovskite PV cells\cite{Yang2017,Kim2017}.\\
The study of phonon-assisted photon absorption from first-principles is notoriously difficult as it involves a double sum over fine grids of k points and complex two excitation processes. Therefore state-of-the-art DFT calculations of phonon-assisted absorption has so far been limited to bulk crystals where the supercell contains only a few atoms\cite{Noffsinger2012}. Recently, Zacharias and Guistino\cite{Zacharias2016} introduced a very efficient method for including phonon induced absorption processes using a single super cell calculation in which the atoms are displaced away from their equilibrium positions. We recently adopted this Special Thermal Displacement (STD) approach to study electron transport in silicon systems with over 1000 atoms including electron-phonon coupling within the DFT-NEGF formalism\cite{Gunst2017}.\\
In this paper we apply STD to calculations of the first order photocurrent in a 19.6\,nm silicon \textit{p-n} junction from DFT-NEGF. In this way we are able to capture the phonon-assisted absorption over the indirect bandgap of silicon and study directly the effect of temperature on the performance of the device. In the following section we summarize the methodology used to calculate the photocurrent using the first order Born approximation and the inclusion of EPC through the STD approach. Exhaustive derivations of the important equations can be found in previous publications\cite{Gunst2017,Zacharias2016,Henrickson2002,Zhang2014,Chen2012}.

\section{Method}
We calculate the photocurrent as a first order perturbation to the electronic system caused by the interaction with a weak electromagnetic field. The electron-photon interaction is given by the hamiltonian
\begin{align}
H'= \frac{e}{m_0} \mathbf{A} \cdot \mathbf{P}
\label{hamiltonian}
\end{align}
where $\mathbf{A}$ is the vector potential and $\mathbf{P}$ is the momentum operator.
Assuming a single-mode monochromatic light source we have\cite{Henrickson2002}
\begin{align}
\mathbf{A} = \mathbf{e}\left( \frac{\hbar\sqrt{\tilde{\mu}_r\tilde{\epsilon}_r}}{2N\omega \tilde{\epsilon} c}F \right)^{\frac{1}{2}}
(be^{-i\omega t}+b^{\dagger}e^{i\omega t}),
\end{align}
where $\tilde{\mu}_r$ is the relative permeability, $\tilde{\epsilon}_r$ is the relative permitivity, $\tilde{\epsilon}$ is the permitivity, $\omega$ is the frequency of the light, $F$ is the photon flux, $N$ is the number of photons, $b^\dagger$ and $b$ are the bosonic creation and annihilation operators and $\mathbf{e}$ is a unit vector giving the polarization of the light. \\
Using the standard Meir-Wingreen formula and including only first order terms in $F$ we arrive at a Fermi's golden rule like expression for the current into lead $\alpha\in L,R$ due to absorption of photons\cite{Rivas2001,Chen2012}
\begin{align}
I_{\alpha} =& \frac{e}{h} \int ^{\infty} _{-\infty} \sum_{\beta=L,R}[1-f_{\alpha}(E)] f_{\beta}(E-\hbar\omega) T^-_{\alpha,\beta}(E) \nonumber \\
& \qquad - f_{\alpha}(E)[1-f_{\beta}(E+\hbar\omega)] T^+_{\alpha,\beta}(E) dE,\\
T^-_{\alpha,\beta}(E) =& N\text{Tr}\{ M^\dagger \tilde{A}_{\alpha}(E)MA_{\beta}(E-\hbar\omega) \}, \\
T^+_{\alpha,\beta}(E) =& N\text{Tr}\{ M\tilde{A}_{\alpha}(E)M^\dagger A_{\beta}(E+\hbar\omega) \}
\end{align}
where $f_{\alpha}$ is the distribution function of lead $\alpha$, $A_{\alpha}=G\Gamma_{\alpha}G^\dagger$ is the spectral function of lead $\alpha$, $\tilde{A}_{\alpha}=G^\dagger\Gamma_{\alpha}G$ is the time reversed spectral function of lead $\alpha$ and the electron-photon coupling matrix is
\begin{align}
M_{ml}=\frac{e}{m_0}\left( \frac{\hbar\sqrt{\tilde{\mu}_r\tilde{\epsilon}_r}}{2N\omega \tilde{\epsilon} c}F \right)^{\frac{1}{2}} \mathbf{e}\cdot \mathbf{P}_{ml}.
\end{align}
The total current is then calculated as $I_{\text{tot}}=I_L - I_R$.\\
The retarded $G$ and advanced $G^\dagger$ Green's functions, spectral broadening of the leads $\Gamma_{\alpha}$ and the momentum operator $\mathbf{P}$ are calculated self-consistently from DFT-NEGF simulations of the silicon \textit{p-n} junction device.\\
The temperature dependent EPC is included through displacements of the atomic positions according to\cite{Gunst2017,Zacharias2016}
\begin{align}
\mathbf{u}_{STD}(T)=\sum _\lambda (-1)^{\lambda -1}\sigma _\lambda(T) \mathbf{e}_\lambda
\label{STD}
\end{align}
where $\mathbf{e}_\lambda$ is the eigenvector of phonon mode $\lambda$ and the Gaussian width $\sigma$ is given by
\begin{align}
\sigma_\lambda(T) = l_\lambda\sqrt{2n_\lambda(T)+1}
\end{align}
where $n_\lambda(T)$ and $l_\lambda$ are the Bose-Einstein occupation and vibrational characteristic length of mode $\lambda$ respectively.
Phonon modes are obtained from a supercell method \cite{Gunst2016}.\\
The configuration displaced according to \eqref{STD} gives the correct thermal average of the Landauer conductance and the optical absorption for sufficiently large systems with many repetitions of the same unit cell \cite{Gunst2017,Zacharias2016}. \\
All DFT-NEGF calculations in this study were performed with the ATK DFT software \cite{ATK,Soler2002,Brandbyge2002} using the SG15-low basis set and 11x11 (21x21) k-points in the electronic structure (transport) calculations. DFT within the local density approximation (LDA) or generalized gradient approximations (GGA) to the exchange correlation potential is known to underestimate the bandgaps in semi-conductors. In order to overcome this bandgap problem, we used the GGA\,+\,1/2 exchange correlation method\cite{FerreiraPRB2008}, which yield accurate band structures for a wide range of materials with the same computational effort as normal GGA calculations. With these parameters we obtain an indirect bandgap of 1.135\,eV at 0\,K.
All phonon calculations were performed using Tersoff potentials\cite{Tersoff1988,Schneider2017}.
\begin{figure}[t!]
\centering%
\includegraphics[width=\columnwidth]{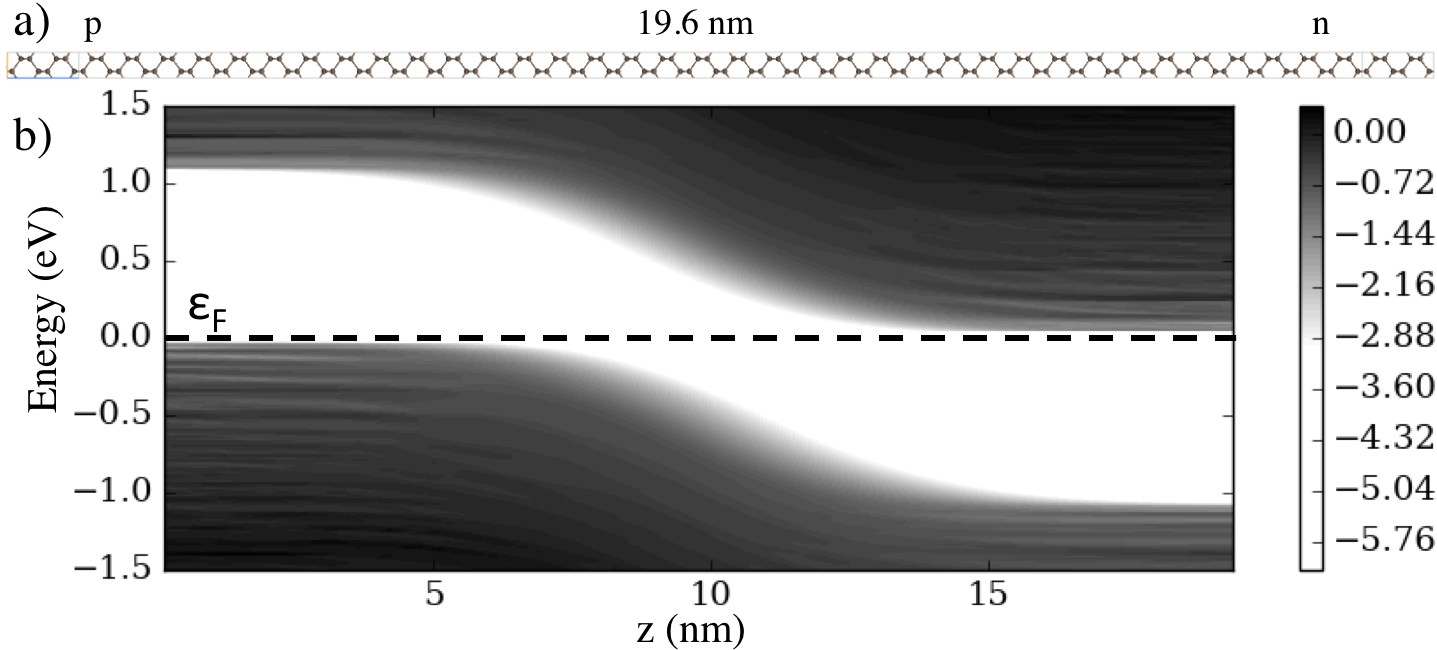}
\caption{a) Structure and cell used in the calculation of the 19.6\,nm silicon \textit{p-n} junction with $n=2.5\times 10^{18}\text{cm}^{-3}$. b) Local density of states along the transport direction of the silicon \textit{p-n} junction on a logarithmic scale.}
\label{fig:ddos}
\end{figure}

\section{Result and Discussion}
\begin{figure}[t!]
\centering%
\includegraphics[width=\columnwidth]{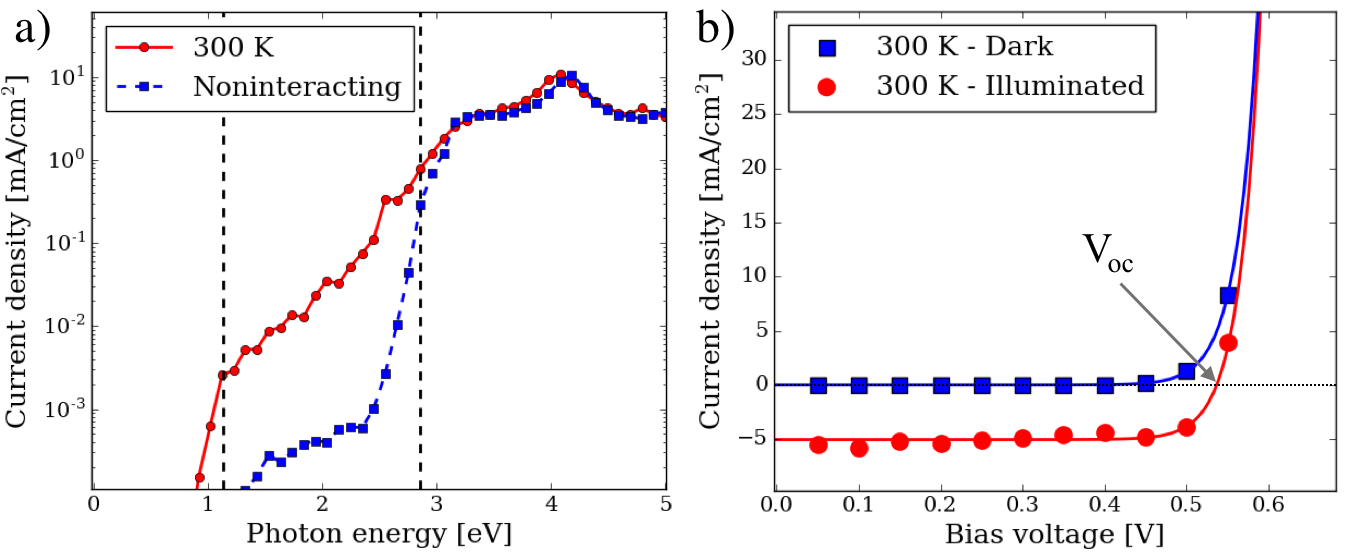}
\caption{a) The calculated photocurrent density for the 0\,K pristine (blue dashed) and STD displaced (red solid) system where $F=1/\text{\AA}^2 s$.
b) The calculated room temperature current density as a function of voltage for the silicon \textit{p-n} junction in the dark $(F=0)$ and under illumination.}
\label{fig:response}
\end{figure}

Fig.\ref{fig:ddos}(a) shows the considered 19.6\,nm silicon \textit{p-n} junction with transport along the [110] direction. The related local density of states is shown in Fig.\ref{fig:ddos}(b) and we can see the typical \textit{p-n} profile along the device with flat bands near the electrode indicating converged screening potentials\cite{Stradi2016,Gunst2017}. Furthermore we see that the calculated bandgap is very close to the one observed in experiments on silicon.\\
In Fig.\ref{fig:response}(a) we compare the photocurrent density calculated for a 0\,K pristine silicon \textit{p-n} junction with that of a 300\,K STD structure for $F=1/\text{\AA}^2 s$. The dashed lines indicate the energies of the indirect ($E_g^{\text{indirect}} = 1.135$\,eV) and direct ($E_g^{\text{direct}} = 2.853$\,eV) bandgaps of the bulk silicon structure using the same calculational settings. Firstly we see that the obtained bandgaps agree well with experimental values verifying our use of GGA\,+\,1/2 exchange-correlation. Secondly we see that the inclusion of EPC through the STD results in an increase in photocurrent from the indirect transition of about 2 orders of magnitude.
The photocurrent as a function of photon energy at 300\,K compares well to previous calculations of the absorption coefficients in bulk silicon where EPC was included in the same way\cite{Zacharias2016}. Temperature effects on the electronic structure are included in the STD method and indeed we see a finite photocurrent at photon energies below the bulk bandgap corresponding to a reduction in the bandgap with temperature.
In addition to the finite temperature renormalization of the band gap, we obtain the actual photocurrent of the transport setup assisted by band-to-band tunneling and including the device potential at finite bias.
Unlike calculations on bulk silicon, we see here a finite contribution to the photocurrent coming from the indirect transition even without EPC. This we trace back to symmetry breaking by the device potential due to the doping profile and finite bias. In simulations based on the bulk silicon band structure such transitions would be prohibited by the opposite symmetry of conductance and valence states if not assisted by phonons. The effects of phonon-assisted tunneling, temperature renormalization and the device potential are all seen to play an important role in the quantitative photocurrent device characteristics.
Regarding below gap transistions, a similar result was seen previously when studying theoretically the phonon assisted tunneling into graphene in a scanning tunneling spectroscopy setup\cite{Palgaard2015}. Here a finite but strongly suppressed tunneling into the Dirac point of graphene is seen even below the threshold voltage of the phonon opening the inelastic channel.
In order to generate an IV curve we need to calculate the total photocurrent under sunlight illumination. To that end we used the AM 1.5 reference spectrum and integrated the spectral current densities for a certain applied bias.\\

\begin{figure}[t!]
\centering%
\includegraphics[width=\columnwidth]{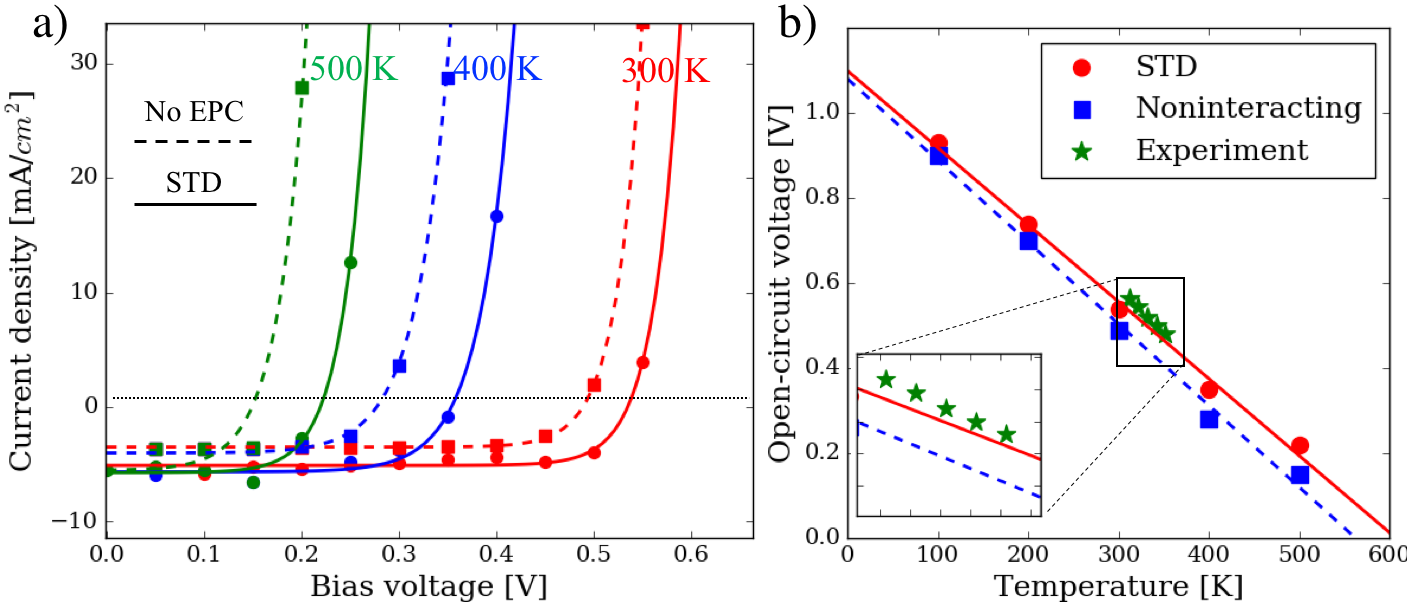}
\caption{a) Calculated IV curves with (solid) and without (dashed) EPC for different temperatures. b) Calculated open-circuit voltage as a function of temperature with linear fits. The green star markers are experimental measurements from [\onlinecite{Huang2011}]. Inset shows a close-up near the experimental measurements.}
\label{fig:opencircuit}
\end{figure}

The IV-curve of the silicon \textit{p-n} junction is shown in Fig.\ref{fig:response}(b) with and without the addition of photocurrent. A least squares fit of the calculated datapoints to the usual expression for the current of a diode under illumination $I=I_{ph} + I_0(\exp(\frac{qV}{nk_bT})-1)$ is added. The result looks very much like what is expected\cite{Wurfel2005} with a photocurrent being a nearly constant contribution at all applied biases. The applied bias voltage where the illuminated IV-curve crosses the zero current density line and no current is generated is known as the open-circuit voltage ($V_{oc}$) and is a measurable parameter used to benchmark solar cell performance. For crystalline silicon PV cells the open-circuit voltage is known to be in the range 0.55-0.60\,V \cite{Huang2011,Loper2012,Chander2015} at room temperature in good agreement with the 0.54\,V obtained from our calculation. The short-circuit current, obtained at zero applied bias, varies a lot depending on the quality and device geometry of the measured solar cell making a direct comparison with our results difficult. The obtained 5\,mAcm$^{-2}$ is however aligned with typical numbers published in the literature\cite{Loper2012,Chander2015}.\\
To analyze the direct impact of EPC on the device we plot in Fig.\ref{fig:opencircuit}(a) the IV-curve under illumination for the perfectly symmetric (noninteracting) silicon system together with that of the system where atomic positions are displaced according to STD \eqref{STD} for different temperatures. Without EPC the short-circuit current density is underestimated by about 25$\%$.
Importantly, the open-circuit voltage is significantly underestimated without EPC and one obtains $V_{oc} = 0.49$\,V at $300$\,K. The open-circuit voltage of a PV device is an important performance indicator that defines its quality and is typically very well controlled, so a $\sim 10\%$ loss at room temperature is substantial. This highlights the importance of including phonon effects in device calculations of indirect semiconductor PV devices.\\

We will now analyze the temperature dependence of the device characteristics which can often be extracted from experiments.
The short-circuit current is largely constant while the open-circuit voltage is degraded for higher temperatures. Comparing with the results where EPC is ignored (dashed lines) we see that the open-circuit voltage is systematically underestimated and that the error is larger at high temperature. It is not surprising that the inclusion of EPC is more important at higher temperatures where the phonon population is higher. The short-circuit current is also constant for the case without EPC, but it is too low at all temperatures.
Using the results shown in Fig.\ref{fig:opencircuit}(a) we can extract the open-circuit voltage as a function of temperature, which is often measured in experiments on solar cell devices. The open-circuit voltage as a function of temperature with and without including EPC is shown in Fig.\ref{fig:opencircuit}(b). A linear fit was performed on both datasets using least squares fitting and the best fit was plotted alongside the datapoints. For both cases we get the expected linear temperature dependence. The open circuit voltage extrapolated to the $T=0$\,K point is often used in experiments to extract the activation energy of the dominant recombination path. Here we get 1.1\,V\,$\pm$\,0.025\,V with EPC and 1.08\,V\,$\pm$\,0.029\,V without EPC both slightly below our calculated bandgap.
In Fig.\ref{fig:opencircuit}(b) we also compare the calculated results with experiments carried out by Huang {\it et al.}\cite{Huang2011}, where the open-circuit voltage of a crystalline silicon PV module was measured under simulated solar irradiation while controlling the cell temperature.
The calculated results including EPC agree nicely with the experimental measurements. The calculated open-circuit voltage without including EPC are much lower than the experimental values at all temperatures. Experimental measurements of $V_{oc}$ performed at temperatures in the range 100-300\,K under 1.1\,Suns illumination\cite{Loper2012} also agree nicely with the values calculated including EPC.\\

\begin{figure}[t!]
\centering%
\includegraphics[width=\columnwidth]{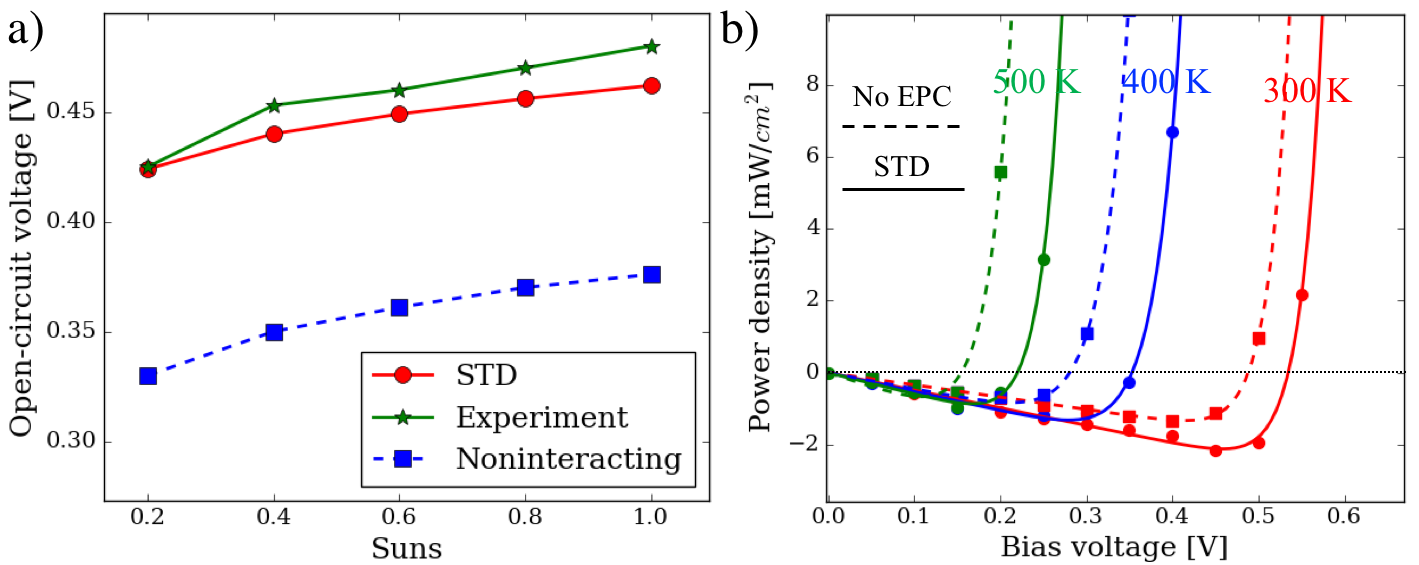}
\caption{a) Open-circuit voltage as a function of light intensity at 352\,K. The calculational results including EPC were extrapolated from the fitted line in Fig.\ref{fig:opencircuit}(b) since we did not perform calculations at the exact temperature measured in \cite{Huang2011}. b) Calculated power density as a function of applied voltage with (solid) and without (dashed) EPC for different temperatures.}
\label{fig:power}
\end{figure}
Lastly, we will analyze the trends with light intensity.
In Fig.\ref{fig:power}(a) we show the calculated open-circuit voltage for different intensities of the light source at a temperature of $352$\,K and compare with experimental values from Huang {\it et al.}\cite{Huang2011}.
Again we see that the results where EPC is included through STD agree nicely with experimental measurements. The best agreement is seen at 0.2\,Suns which is expected given the assumption of a weak field going into Eq.~\eqref{hamiltonian}. On the other hand the open-circuit voltages calculated while neglecting EPC do not agree with experimental values. In fact, even the results calculated at 1\,Sun are below experimental values where only an intensity of 0.2\,Suns are used. This underlines the pivotal role played by EPC in PV devices.
Fig.\ref{fig:power}(b) shows the generated power density for the \textit{p-n} junction with and without including EPC. We see that the maximum power and maximum power point are both underestimated in the noninteracting case.

\section{Conclusion}
We have presented a computationally cheap method to calculate the phonon-assisted photocurrent in large-scale devices from first-principles. Previous studies of phonon-assisted optical absorption using state of the art methods have been limited to bulk systems of high symmetry where only a handful of atoms are considered due to the computational cost. Here we study a 19.6\,nm long silicon \textit{p-n} junction under working conditions with an applied bias. The calculated current density agrees with previous studies of temperature renormalization in the optical absorption of silicon using a similar theoretical approach\cite{Zacharias2016}.
Our results agree nicely with experiments both for values of the open-circuit voltages and trends in how it scales with temperature and light intensity.
The phonon interaction has a significant impact on the device characteristics highlighting the need for photocurrent transport calculations including phonon coupling when considering devices using indirect semiconductors as the absorber material.
The combination of device simulations with photon coupling as well as phonon coupling, through special thermal displacements, gives an appealing way forward in the hard problem of combined light-matter interaction, phonon-assisted tunneling, temperature renormalization and the nonequilibrium device potential in quantitative first-principles simulations.

\begin{acknowledgments}
This work is partly funded by the Innovation Fund Denmark (IFD) under File No. 5016-00102. We thank Andrea Crovetto for his input on this work.
\end{acknowledgments}

\end{document}